\def\kms{\ifmmode{\,\hbox{km}\,s^{-1}}\else {\rm\,km\,s$^{-1}$}\fi}

\def\msun{{\rm\,M_\odot}}
\def\lsun{{\rm\,L_\odot}}

\def\kmsm{{\rm\,km\,s^{-1}\,Mpc^{-1}}}

\def\hmpc{\ifmmode{h^{-1}\,\hbox{Mpc}}\else{$h^{-1}$\thinspace Mpc}\fi}

\def\et{{\it et~al.}~}

\documentclass[12pt,preprint]{aastex}
\begin{document}
\relax
\tolerance 9000

\title
{Globular Clusters at High Redshift}

\author{R.~G.~Carlberg}

\affil{Department of Astronomy \& Astrophysics}
\affil{University of Toronto, Toronto ON, M5S~3H8 Canada}
\email{carlberg@astro.utoronto.ca}


\begin{abstract}
Globular clusters will be present at high redshifts, near the very
beginning of the galaxy formation process. Stellar evolution ensures
that they will be much more luminous than today. We show that the
redshift distribution at nano-Jansky levels should be very broad,
extending up to the redshift of formation.  A bracketing range of
choices for the redshift of formation, spectral energy evolution
models and population density evolution, leads to the conclusion that
the sky densities should be around $10^7$ per square degree at 1~nJy
($m_{AB}=31.4$ mag) in bands around 4 microns. Such high sky densities
begin to present a confusion problem at these wavelengths to
diffraction limited 6m class telescopes.  These star-like, low
metallicity, clusters will be a significant foreground population for
``first light'' object searches. On the other hand they are an
exceptionally interesting ``second light'' population in their own
right. Depending on the details of galaxy assembly, the clusters will
have a noticeable cross-correlation with galaxies on scales of about
20 arcsec, or less, depending on the details of the buildup of galaxy
assembly after globular cluster formation. High redshift globular
clusters will be an accessible, direct, probe of the earliest stages
of the formation of galaxies and the buildup of metals in the
universe.
\end{abstract}


\clearpage
\section{Introduction}

Globular clusters contain some of the oldest stars in the universe and
have long been vital clues to the earliest phases of the star
formation in galaxies \citep{searle+zinn,harris_ext,cmw}. In our own
galaxy the known globular clusters are very old \citep{donv} but there
is evidence that they can form at lower redshift in suitably extreme
conditions, generally associated with merging galaxies
\citep{zepf,zf,ashman+zepf,cen,larsenetal}.  The great age and
low metallicity of globular cluster systems indicates that they should
be present at very high redshifts and predate the bulk of their
eventual host galaxies' stars.

The exciting prospect is that direct studies of globular cluster
formation and evolution will soon become possible.  The next
generation of optical-infrared telescopes on the ground and in space
will have the capability to detect objects at the nano-Jansky
level. An estimate of the faint number counts in the optical was
undertaken for HST \citep{vdb} but we concentrate on the IR where the
redshifts and rise in numbers is much more dramatic.  In the 2 to 5
micron bands, the combination of large k-corrections and substantial
stellar brightening raises the fluxes from high redshift clusters into
the range of 29-32 AB mag. These nanojansky flux levels are within the
capabilities expected of future telescopes.

Today's globular globular clusters are likely the survivors of a
larger population present at the various times of formation
\citep{fall+rees,fall+zhang}. If their co-moving density increases by
an order of magnitude over those at low redshift then the globular
clusters are likely to appear with numbers at a given flux level that
are comparable to sub-galactic mass dark matter halos which are the
sites of the ``first stars'' \citep{hugh+martin,haiman+loeb,HAM}.

This paper calculates the magnitude limited distribution of the
expected numbers, $n(m)$, the redshift distribution, $n(z|m)$, and
estimates the angular clustering properties of the globular cluster
population relative to their host galaxies.  The predictions are made
for filter pass bands sufficiently red that Lyman $\alpha$ trough
absorption will not normally be an issue. In the next section we
describe the calculation of the co-moving number density of globular
clusters (GC) as a function of redshift for different evolutionary
assumptions. In Section 3 we present the results of the number
calculations. Section 4 considers the apparent sky clustering of the
distant GC. We conclude with a discussion of the opportunities and
complications that this population presents. The calculations are
presented in a cosmology for which $H_0=70
\kmsm$, $\Omega_M=0.3$, $\Omega_\Lambda=0.7$.

\section{An Evolving Globular Cluster Luminosity Function}

The expected sky density of GCs at magnitude $m$ and redshift $z$
depends on the product of the cosmological volume element and their
luminosity function, $\phi_{GC}(L,z)$, integrated with the volume
element along the line of sight.  The luminosity function has three
sources of evolution. First, it is generally accepted that galactic
tidal fields and stellar dynamical ``shocks'' erode a more numerous
high redshift GC population into the remnant population we see
today. We use the results of a relatively secure theoretical analysis
of the evolution of the population, but also show results for a
non-evolving distribution. Second, as the stellar population becomes
younger with increasing redshift its spectral energy distribution
changes. Third, the GCs form at some high, but as yet poorly
determined, redshift. Our approach to each of these evolutionary terms
along with the normalization to the present day globular cluster
population is discussed in the following section.

\subsection{An Evolving Mass Distribution}

Recently Fall \& Zhang (2001, hereafter referred to as FZ) have
discussed a generalized dynamical model for the evolution of the mass
distribution of GCs. They find that within 1-2 Gyr of origin, a wide
range of initial GC mass distribution assumes a characteristic form
which then evolves in a nearly self-similar way. At small mass, all
clusters (in the same tidal field) go to zero mass at the same rate
due to two-body relaxation driven evaporation. At high mass,
gravitational shocks impose a characteristic maximum mass above which
there is a rapid cutoff of numbers. The continual depletion of GCs
implies that over a Hubble time about 10\% of the initial cluster
population survives, under the assumptions that the system is not
replenished and that the galactic potential does not change. FZ have
kindly made their results for the evolution of the globular cluster
mass distribution available for use in this paper. Specifically we use
the differential number of GCs at mass $M$ at time $t$, $n_{GC}(M,t)\,
dM$, which FZ present in their Figure 3.

It should be noted that the FZ model {\it predicts} the mass of the
peak.  To test this aspect of the models FZ have put the Milky Way
globular clusters on a mass scale using $M/L_V=3$, as is appropriate
for an old, metal poor stellar population. The agreement between model
and observation is impressively good.  The extensive testing of FZ
shows that the results should not change much with galaxy mass or
galaxy type.  

The FZ model results are specified at times of 0.01, 1.5, 3, 6 and 12
Gyr.  We use a double spline function in the variable $M$ and
$\log{t}$ to interpolate to other times. We extrapolate slightly
beyond their 12 Gyr model to the age of 13.4 Gyr age of our
cosmology. The minimum age of their models is a 0.01 Gyr, where the
``formation distribution'' is close to a power law in mass. We will
show the sensitivity of our results to the GC number evolution.

\subsection{The Redshift Dependent Luminosity Function}

We require the redshift dependent globular cluster luminosity function,
$\phi_{GC}(L_{\bar{\lambda}},z)$, where $L_{\bar{\lambda}}$ is the
observed luminosity in some filter band centered around
$\bar{\lambda}$.  The conversion from $n_{GC}(M)$ to $\phi_{GC}(L)$ is made
using a spectral synthesis model which gives the entire spectral
energy distribution, $F_\lambda$, as a function of model age for given
metallicity and star formation history.  We use the PEGASE.2 code
\citep{pegase} to calculate $\ell_{\bar{\lambda}}$, 
the observed frame luminosity per unit mass in the filter band
$\bar{\lambda}$,

\begin{equation}
\ell_{\bar{\lambda}} = {{\int_0^\infty T(\lambda)F_\lambda((1+z)\lambda) 
	\,d\lambda}
	\over{(1+z)\int_0^\infty T(\lambda) \,d\lambda}},
\label{eq:ml}
\end{equation}
where $F_\lambda[(1+z)\lambda]/(1+z)$ is the model's mass normalized
absolute flux in the observed frame and $T(\lambda)$ is the filter
transmission function.  Noting that the photon redshift and the time
dilation are included in Eq.~\ref{eq:ml}, 
an object of mass $M$ gives an observed flux in the
$\bar{\lambda}$ filter of
\begin{equation}
f_{\bar{\lambda}} =  {{M\ell_{\bar{\lambda}}}\over{4\pi r^2(z)}},
\label{eq:flux}
\end{equation}
where $r(z)$ is the co-moving distance in the adopted cosmology.  The
observed flux is converted to magnitudes using the definition
$m_{\bar{\lambda}}\equiv -2.5\log_{10}(f_{\bar{\lambda}}) + C$, where
$C$ is 31.4 AB magnitudes at 1 nano-Jansky. 

\subsection{Normalizing the Luminosity Function}

The luminosity functions of the GC systems of the Milky Way and more
than 50 nearby galaxies have been studied
\citep{harris_ext,harris_mw}.  A single galaxy's GC luminosity
function is conventionally described as a Gaussian (in absolute
magnitude, hence a lognormal distribution in luminosity) centered at
$\langle M_V
\rangle= -7.27+5\log_{10}{(H_0/75)}$ mag with a dispersion of about
1.2 magnitudes.  Although a more complex function, the FZ mass model
appears to describe the data at least as well as a Gaussian. Moreover,
it is based on a dynamical theory that allows its history to be
predicted. To use the FZ function in our calculation we need to fix
the volume normalization and we will also introduce a small shift in
the $M/L$ value.

\subsubsection{Mass-Luminosity Normalization}

The mass normalization of the FZ models is determined by the dynamics
of the GCs within the model galaxy. Although fairly insensitive to
variations in the potential, the mass function does shift slightly
depending on the specific galactic potential.  Here we need the
luminosity function typical of a mix of galaxies. We adopt the
functional form of the FZ mass function and could adopt their
$M/L_V=3$ value, however we prefer to make a small adjustment to
provide an alternate match to the observational data.  The FZ mass
function is a power law on the low mass side and much steeper than a
Gaussian on the high mass side. Here we chose an $M/L_V$ value that
brings the mean luminosities of the FZ distribution to the mean of the
Gaussian fits. A numerical integration finds that $M_V=-7.27$ mag
should be identified as $\log{M/\msun}= 5.36$, which implies an
$M/L_V=3.3\msun/\lsun$.  This small $M/L$ change is well within the
uncertainty of stellar population modeling.  In particular, $M/L_V$ is
$2.1\msun/\lsun$ for the $Z=0.1$ solar PEGASE models we compute at an
age of 13.4 Gyr. Our normalization effectively raises the $M/L$ values
of the PEGASE models by a multiplicative factor of 1.57.

The V band luminosity is widely used to describe low redshift
clusters. However, it is beneficial for the accuracy of our
application to high redshift galaxies main application to use K band
luminosities. Furthermore, the GC population is most closely connected
to the old stellar population which is most accurately measured at low
redshift by K band luminosities.  The conversion from V to K must use
the IR colors of a population with a metal abundance of about
one-tenth solar, $V-K=2.93 +0.5 Z/Z_\odot$ mag \citep{vk_colors}.  The
$Z=0.1Z_\odot$ PEGASE models find $V-K=2.4$ mag at 13.4 Gyr which is
in essentially exact agreement with the observational relation.  Using
this color we find that the mean peak K band luminosity for GCs is
$\langle M_K \rangle =-9.70+5\log_{10}{(H_0/75)}$. This value is
converted to the flux based AB magnitude system with the addition of
$AB(K)=+1.88$ mag.

\subsubsection{Number Density Normalization}

The mean co-moving density of GCs for a single galaxy is modeled as
being directly proportional to its luminosity,
\citep{S,harris_ext},
\begin{equation}
S_N= N_t 10^{-0.4(M_V+15)},
\label{eq:SV}
\end{equation}
where $M_V$ is the galaxy's absolute magnitude in the V band and $N_t$
is the total number in a Gaussian luminosity function. The $S_N$
relation has significant variations with Hubble type and possibly
environment but appears to be accurate in the mean \citep{harris_ext}.
Since GC are most clearly associated with old stellar light it is
natural to use a K band luminosity function, in which case,
\begin{equation}
S_N= N_t 10^{-0.4(M_K+17.9)},
\label{eq:SK}
\end{equation}
where a galaxy with solar metallicity has $V-K\simeq 2.9$ mag
\citep{vk_colors}. 
Adopting the Gardner \et\ (1993) luminosity
function (for our purposes, similar to the recent Cole et al. 2001
results), for which $M_\ast(K) = -23.1$ mag, we find that in
the K band the number of globular clusters around a galaxy rises
linearly with luminosity,
\begin{equation}
N_t(L_K) =120S_N {L_K\over {L_\ast(K)}}. 
\label{eq:nk}
\end{equation}

To convert the normalization from globular cluster per galaxy to 
a volume normalization we use the luminosity density of galaxy light 
in the K band, $j(K)$. The normalizing constant for the GC luminosity
function is defined such that the integral over all GC luminosities
must be equal to the mean number of GC expected for the mean amount of
galaxy light in that volume. That is,
\begin{equation}
\int_0^\infty \phi_{GC}(L_{\bar{\lambda}},z=0)\, dL
	 = 120 S_N {j(K)\over L_\ast(K)}.
\label{eq:norm}
\end{equation}
For Gardner's (1993) $\alpha=-1$ Schechter luminosity function fit
$j(K)=\phi_\ast(K) L_\ast(K)$, where
$\phi_\ast(K)=0.0166h^{-3}$~Mpc$^{-3}$. Note that the dependence on
$L_\ast(K)$ cancels in Eq.~\ref{eq:norm}.  Then the co-moving volume
density is GCs is $n_{GC}(0)=2.0h^{-3}S_N$~Mpc$^{-3}$, where
$h=H_0/100$.  We adopt $S_N=2$ as a reasonable and somewhat
conservative value, given that the bulk of the K light emerges from
relatively luminous early type galaxies. Figure~4 and Table~3 of
Harris (1991) might suggest a value of about 3 for the early type
galaxies, with evidence that strongly clustered early type galaxies
have higher $S_N$. Of course the origin of these effects may well be
directly visible in the future.  The outcomes is that our complete GC
luminosity function is,
\begin{equation}
\phi_{GC}(L_{\bar{\lambda}},z) \,dL_{\bar{\lambda}} = 
	n_{GC}(0)
	n_{GC}[L_{\bar{\lambda}}/\ell_{\bar{\lambda}}(t),t(z)] 
	\,dL_{\bar{\lambda}}.
\end{equation}

The redshift distribution per unit sky area of GC at a given flux
level, is simply
\begin{equation}
n(z|f_{\bar{\lambda}}) \, d\ln{f_{\bar{\lambda}}} = \int_0^\infty \phi_{GC}(4\pi
	r^2(z)f_{\bar{\lambda}},z) {{dV}\over {dz}}\, dz\, d\ln{f_{\bar{\lambda}}}.
\label{eq:sky}
\end{equation}
where $dV/dz$ is the volume element within the model
cosmology. Integrating over the redshift distribution gives the
number-magnitude relation In practice, these calculations are done
using magnitudes rather than fluxes.

\section{Counts and Distributions}

With the modeling apparatus in hand we first pause to show the low
redshift $n(z)$ at $m_R(AB)=25, 26, 27$ and $28$ mag in
Figure~\ref{fig:nzR}.  The total sky densities of the FZ model at
these depths are 87, 321, 1,260 and 4,840 per square degree per
magnitude, respectively.  These GCs will be clearly associated with
relatively bright galaxies, typically about $m_R(AB)=13-17$ mag, with
the redshift distributions shown.

\subsection{Number Redshift Distributions}

Precisely how globular clusters form is, at this time, unknown.  Part
of the point of this paper is that the plausible formation redshifts
for the bulk of GCs will shortly come within reach of telescopes. To
try to bracket the situation, we examine a number of somewhat extreme
alternative models and look at the effects of co-ordinated bursts in a
galaxy. To examine the importance of ``luminosity spikes'' at the time
of formation, we use (arbitrarily, for the purpose of illustration) 10
bursts of star formation of duration 10~Myr spread over the 0 to 1 Gyr
time interval.  The formation age of all of the GC is put at 0.5
Gyr. The results are shown in Figure~\ref{fig:nzL}. Such bursts would
only effect the counts in an area small enough that only a few dozen
actively GC forming galaxies were present.

In Figure~\ref{fig:nzL_alt} we use the same set of ten star formation
bursts but shifted in time to the 1 to 2 Gyr time interval with a
uniform formation age of 1 Gyr. Clearly the bursts of star formation
produce spikes in the redshift distribution but those effects quickly
die away. It could be that the earliest phases of globular cluster
formation are cloaked in dust which later disperses, following an
age-extinction relation \citep{shapley}.  In that case the high
luminosity peaks will be a briefly obscured phase in the life of
GCs. However the Figures show that if those short-lived bright spikes
are removed, neither the counts or redshift distribution will be
greatly altered.

The difference between the no-evolution and FZ density evolution
models are small at redshifts below about three, under the assumption
that most globular clusters were formed at redshifts greater than
three. The differences would be much larger if significant globular
cluster formation continued to much lower redshift.  The numbers
predicted with a non-evolving mass model are nearly a full decade
below the evolving model beyond redshifts of five. Since density and
luminosity evolution are independent in these models the same
difference applies to all formation histories.

\subsection{Number Magnitude Relations}

The number-magnitude relation is shown in Figure~\ref{fig:nm} in the
V, R, J, K, L and M bands (spanning roughly 0.5 to 5 microns) for our
model with ten, bursts of 10\% of the mass, spread between 0 and
1~Gyr. There are two effects. At low redshift the counts rise with
increasing magnitude at a rate governed by the volume element,
enhanced by k-corrections in an old, low metallicity, population.
Although not shown, at 1~nJy in the V band the $n(z)$ peaks at about
redshift 0.3. At 1~nJy in the R band $n(z)$ peaks at about redshift
0.5, with a few of the actively star-forming $z=5$ clusters in
formation being visible.  Clearly deep optical band observations are
not the ideal way to probe the formation epoch. The character of the
redshift distribution changes in the infrared bands as the peak of the
spectrum is redshifted into them. The combination of k-correction and
luminosity evolution causes the counts in the redder bands to rapidly
climb to several million per square degree.  As Figure~\ref{fig:nm}
shows, the counts are steeper than Euclidean near 30 AB mag in the IR
bands.

The predicted counts for a wide range of model star formation
histories are shown in Figure~\ref{fig:nmE}.  We display the L band
counts for models having ten bursts of 10\% of the final mass star
formation extending over 10~Myr in the 0-1 Gyr interval (triangles),
the 1-2 Gyr interval (diamonds), and exponential models with $\tau=1$
pentagons), 2 (heptagons) and 4 Gyr (hexagons) for both evolving and
non-evolving mass function. The number-magnitude relation shows
substantial model dependencies beginning at about 1~nJy ($m_{AB}=31.4$
mag). However, the result that the L band counts will be around $10^8$
per mag per square degree at 0.2~nJy is reasonably robust. It does not
depend a lot on star formation, internal internal dust shrouding in
the early phases, and is not unduly sensitive to the exact amount of
GC density evolution. The biggest potential over-prediction of numbers
is if GC initially form in dusty disk environments which makes them
hard to detect. As long as the clusters become visible within about 2
Gyr of formation the numbers predicted here should be fairly
accurate. The basic prediction that the sky density becomes about
$10^7$ per square degree around 1 nJy is difficult to escape, given
the assumptions about globular cluster origins, evolution and
visibility made in this paper.

\section{Angular Clustering}

Globular clusters are strongly concentrated around their host
galaxies.  At the very low flux levels we are investigating here it is
natural to ask to what degree this clustering will remain evident and
whether the nano-Jansky sky will effectively be covered with a nearly
uniform distribution of GC. A prediction of clustering uses the
results above but requires additional information about the degree to
which galaxies and their GC systems merge into larger and larger
units. Furthermore, the host galaxies may not always be visible at the
highest redshifts considered here, since galaxies are generally
younger and much lower surface brightness than GCs. Hence, the
following estimates of galaxy-GC cross-correlations will be upper
limits, although we do incorporate a model for galaxy merging into our
calculations.

The real space cross-correlation of galaxies and GCs can be derived
from the average radial profile of GCs in their host galaxies. As
shown below the auto-correlation of galaxies makes no significant
contribution at the angles of interest.  The FZ calculations find that
after approximately 1 to 2 Gyr the radial distribution converges to a
stable, nearly power-law form.  An approximate power law fit to the
Milky-Way data of Harris (1996) is,
\begin{equation}
n(r)=3\times 10^2 \left({r\over{10 {\rm kpc}}}\right)^{-3.5} {\rm kpc}^{-3}.
\label{eq:gcr}
\end{equation}
If there is a core in the radial distribution it appears at a radius
of order a few kpc where the GCs become superimposed on significant
galaxy light and hard to find. The Milky Way is probably somewhat less
than $L_\ast$ in luminosity and its GC system, with a total of
$160\pm20$ clusters \citep{harris_ext}, has numbers about 2/3 half of
the 240 expected at $L_\ast$.  To calculate the cross-correlation
function with galaxies we need $\delta(r)= (n(r) -n_0)/n_0$, where
$n_0$ is the mean density.  We normalize these numbers to the volume
average for $L_\ast$ galaxies.  The mean GC density of
$4.0h^{-3}$Mpc$^{-3}$ we derived above becomes a physical density of
$1.2\times 10^{-8}$~kpc$^{-3}$ for $H_0=70$. Consequently we can
re-express Eq.~\ref{eq:gcr}, as the over-density,
\begin{equation}
\delta(r)=2.5\times 10^{10} \left({r\over {10 {\rm kpc}}}\right)^{-3.5}.
\label{eq:delta}
\end{equation}
Converting this to the standard correlation length form and using
co-moving co-ordinates ($H_0=70$),
\begin{equation}
\xi_{gGC}(x) = \left({{9.4 {\rm Mpc}}\over x}\right)^{3.5}
		{L_h\over L_\ast},
\label{eq:xi}
\end{equation}
where we have included the luminosity dependence, with $L_h$ being the
luminosity of the host galaxy.  Note that an alternate description of
this correlation length is $6.6\hmpc$. In this calculation we have
assumed that the low redshift $S$ relationship holds in the earliest
phases of the life of a galaxy which needs to be tested.  An overall
density normalization change has no effect on the correlations since
the mean field density changes at the same rate, leaving $\delta$
invariant. 

GC systems appear to always be associated with more or less virialized
galaxies. They are not part of a clustering hierarchy that extends
into the linear regime. Therefore we describe the over-density
distribution as being fixed in physical co-ordinates. We therefore
multiply Eq.~\ref{eq:xi} by the correlation function
evolution term $(1+z)^\epsilon$. The quantity $\epsilon$ is equal to
$\gamma-3$ for fixed over-density in physical co-ordinates, as is
appropriate here.

Galaxies are assembled over time through the merger process.  A simple
model for the increase of mass $M$ is $dM/dt = {\cal R} (1+z)^{\cal
M}$.  Approximating $1+z=t_0/t$ (as in an empty universe) this
integrates to 
\begin{eqnarray}
         & M_0 - {\cal R}({\cal M}t_0)^{-1} (1+z)^{{\cal M}-1}, 
		&  {\cal M} > 1, \\
M(z) =\Bigg\{  &   &  \\
	& M_0-{\cal R}t_0 \log{(1+z)},~~~~~~~ & {\cal M}=1.  \\
\end{eqnarray}
Note that $M(z)$ goes to zero for finite $z$ for ${\cal
M}>1$.  Reasonable values are ${\cal M}\simeq 1-3$ and ${\cal
R}t_0\simeq 0.2-0.5$ \citep{mergers,cfrs_mg}. The resulting redshift
dependent co-moving correlation function, $\xi_{gGC}(x|z)$, is,
\begin{equation}
\left({{r_0(z)}\over x}\right)^\gamma 
= (1+z)^\epsilon \left( {{r_{00}}\over x(z)}\right)^\gamma
	{M(z)\over M_0}.
\label{eq:xiz}
\end{equation}

The angular correlation function is simply related to the volume
correlation through a projection over redshift,
\begin{equation}
\omega(\theta) = A(\gamma)
\theta^{1-\gamma} N^{-2}
\int n^2(z) \left({{r_0(z)}\over x}\right)^\gamma x  \, {{H(z)}\over{c}} dz,
\end{equation}
where $N=\int n(z)\, dz$,
$A(\gamma)=\Gamma(\onehalf)\Gamma((\gamma-1)/2)/ {\Gamma(\gamma/2)}$,
and $H(z) = H_0
[\Omega_M(1+z)^3+\Omega_R(1+z)^2+\Omega_\Lambda]^{1/2}$, with
$\Omega_M+\Omega_R+\Omega_\Lambda=1$. 

We express the results as an angular correlation $\omega(\theta) =
(\theta_0/\theta)^{\gamma-1}$. We evaluate the integral using the L
band $n(z)$ at 1nJy. For our $r_{00}=6.6\hmpc$ Mpc, $\gamma=3.5$,
$\epsilon=0.5$, we find $\theta_0 = 22$ and $18\arcsec$, with ${\cal
R}t_0=0.5$ for ${\cal M}=1$, and $2$, respectively and 21 and
$17\arcsec$ for ${\cal R}t_0=0.3$ for the same ${\cal M}$.
Correlation angles of $20\arcsec$ correspond to physical distances of
about 100~kpc around redshift three. Therefore the bulk of the GCs
will be clearly associated with their host galaxies. The galaxies, if
they exist and are not obscured, will be some $\sim10-12$ mag brighter
than the GCs, depending on the relative roles of merging and
luminosity evolution. At $K_{AB}\simeq 20-22$ mag galaxies have mean
sky separations of $\sim 20-40\arcsec$, so the sky will be effectively
covered, albeit with a concentration toward galaxies, or, the
still-dark halos that will become the sites of galaxies.

The GC-galaxy cross-correlation calculation ignores the contribution
due to galaxy-galaxy clustering. The same of calculation shows that
the much shallower $\gamma=1.8$ of galaxy clustering the
auto-correlation angle is about 2 arcsec, for $r_{00}=5\hmpc$. For the
GCs their steep cross-correlation with galaxies allows them to rapidly
climb out of the projected distribution, which does not occur for the
galaxy-galaxy correlation. The galaxy-galaxy contribution will only be
visible at about an arc-minute, where the projected clustering
amplitude is only $\sim 0.03$. At angles less than 20\arcsec\ the
contribution is less than 10\%, given our modelling for clustering.
Of course globular cluster formation during merging is a special case.

\section{Discussion and Conclusions}

Globular clusters are, in the main, very old objects, likely formed in
the first quartile of the age of the universe, implying strong
luminosity evolution at high redshift. The combination of
k-corrections and luminosity evolution put the bulk of their energy in
the 3 to 5 micron bands. For a fairly wide range of density and
luminosity evolution models there should be approximately $10^7$ GC
per square degree per magnitude, with a continuing steep rise in
counts.  In the optical bands the counts rise slowly with few GC
appearing beyond redshift one.

Source confusion noise in flux and position measurements increases in
proportion to the density of sources relative to the beam density,
$(\ln{2}/\pi) (D/\lambda)^2$
\citep{scheuer,condon}.  The problems associated with confusion
begin to arise when the source density is about $\sim5$\% of the beam
density. Moreover the strong clustering of GC toward galaxies will
create enhanced confusion in the neighborhood of galaxies.  A
diffraction limited 6m telescope operating at $4\mu$ will have one
source per beam (severe confusion) at a sky density $1.5\times 10^8$
per square degree. For the relatively steep source counts found here
and the high confidence detections that would be of interest to
photometric redshift estimation techniques, the source density below
about 1~nJy presents an issue to be carefully approached. In detail
this problem could be more quantitatively addressed with simulated
observations using the predicted counts.  In a future paper we will
also consider a more detailed model that incorporates dust and
emission line nebula effects and a number of potential astrophysical
complications.

The large density of high redshift GCs is both an opportunity and a
challenge. In as much as GC are key indicators of how the extended low
metallicity stellar halos of galaxies came into being, observations at
nJy flux levels will directly probe their origins. On the other hand,
the sky densities and flux levels are similar to those predicted for
zero metallicity, ``first light'' objects. It will require some care
to distinguish a young cluster of fairly normal stars with strong
ionizing radiation from the unusual zero metallicity stars that are
the first luminous objects. It will be fascinating to understand the
relationship between these two ``early light'' populations.

\acknowledgements

I thank Mike Fall \& Qing Zhang for providing the results of their
mass evolution models.  Mike's comments also improved the presentation
of the results. Chris Pritchet and Sidney van den Bergh provided
inspiration and comments on this subject.  Research support from NSERC
and CIAR are gratefully acknowledged.

\newpage
\newcounter{figi}
\newcommand{\nfig}{\addtocounter{figi}{1}\thefigi}

\figcaption[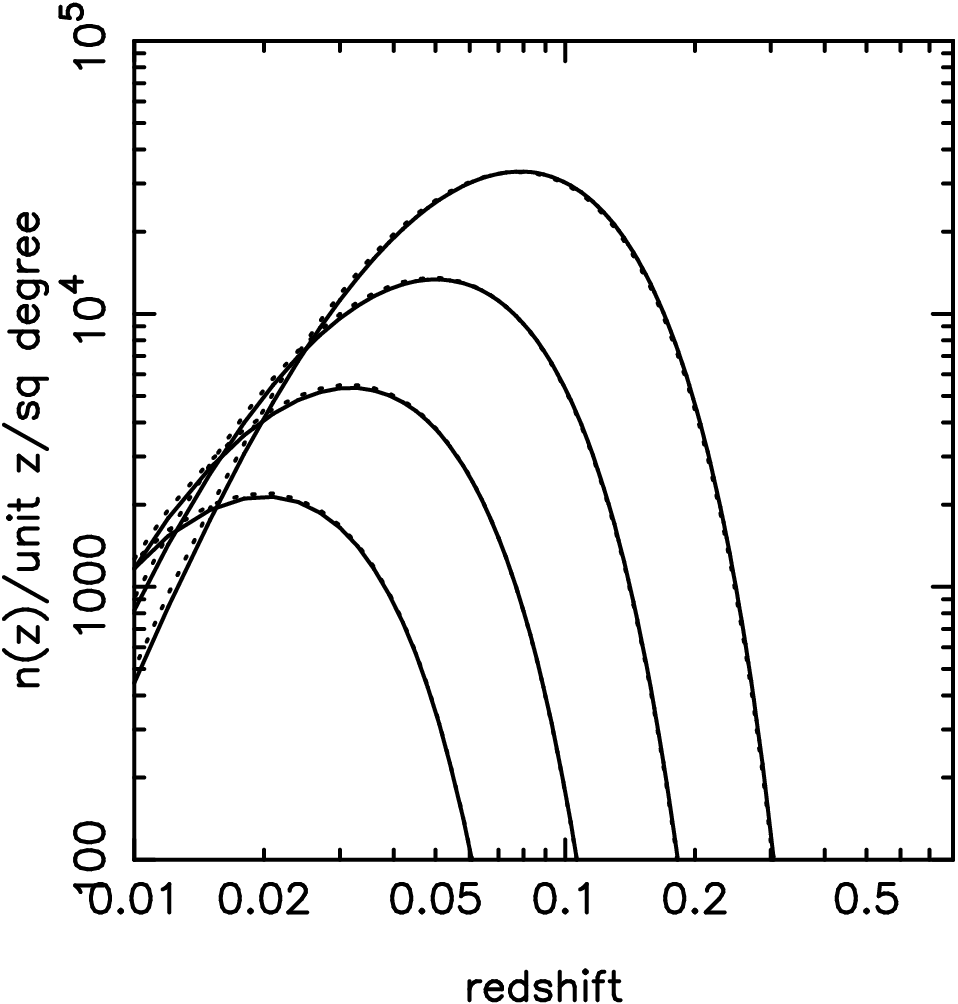]
{The redshift distribution for $m_R(AB)= 25, 26, 27$ and 28 mag,
comparing our evolving luminosity function (solid line) with its
no-evolution form (dotted). 
\label{fig:nzR}}

\figcaption[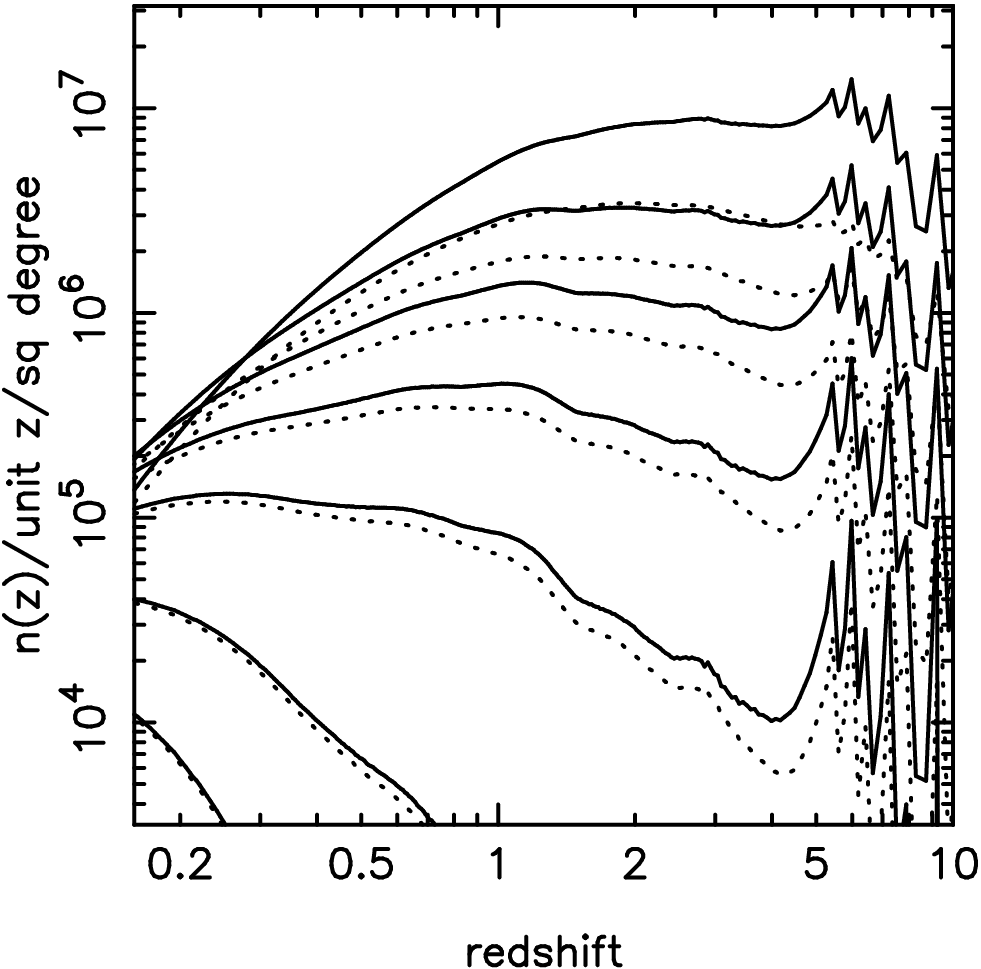]
{The redshift distribution in the L band (around 3.5 microns) where
the GCs form over the time range of zero to one Gigayear in 
10 bursts of 10Myr. The solid line is for the density evolution model the
dotted line is for no density evolution. Curves are presented
for 0.2, 0.5, 1, 2, 4, 10 and 20 nJy, equivalent to 33.15, 32.15,
31.4, 30.65, 29.9, 28.9 and 28.15 AB mag, respectively.
\label{fig:nzL}}

\figcaption[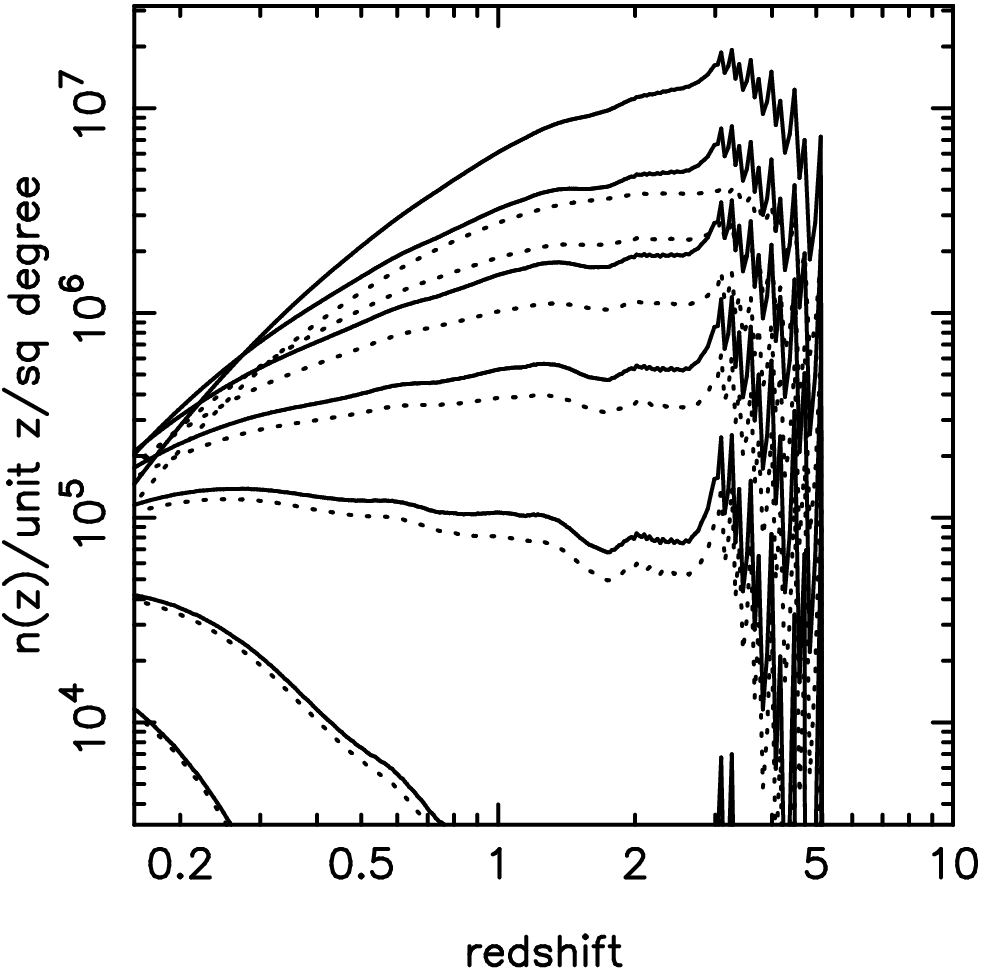]
{The redshift distribution in the L band for 10 bursts of 10Myr of GC
formation over the 1 to 2 Gyr time interval. Line types are as in
Fig.~\ref{fig:nzL}. The same magnitude limits as in Fig~\ref{fig:nzL}
are used.
\label{fig:nzL_alt}}

\figcaption[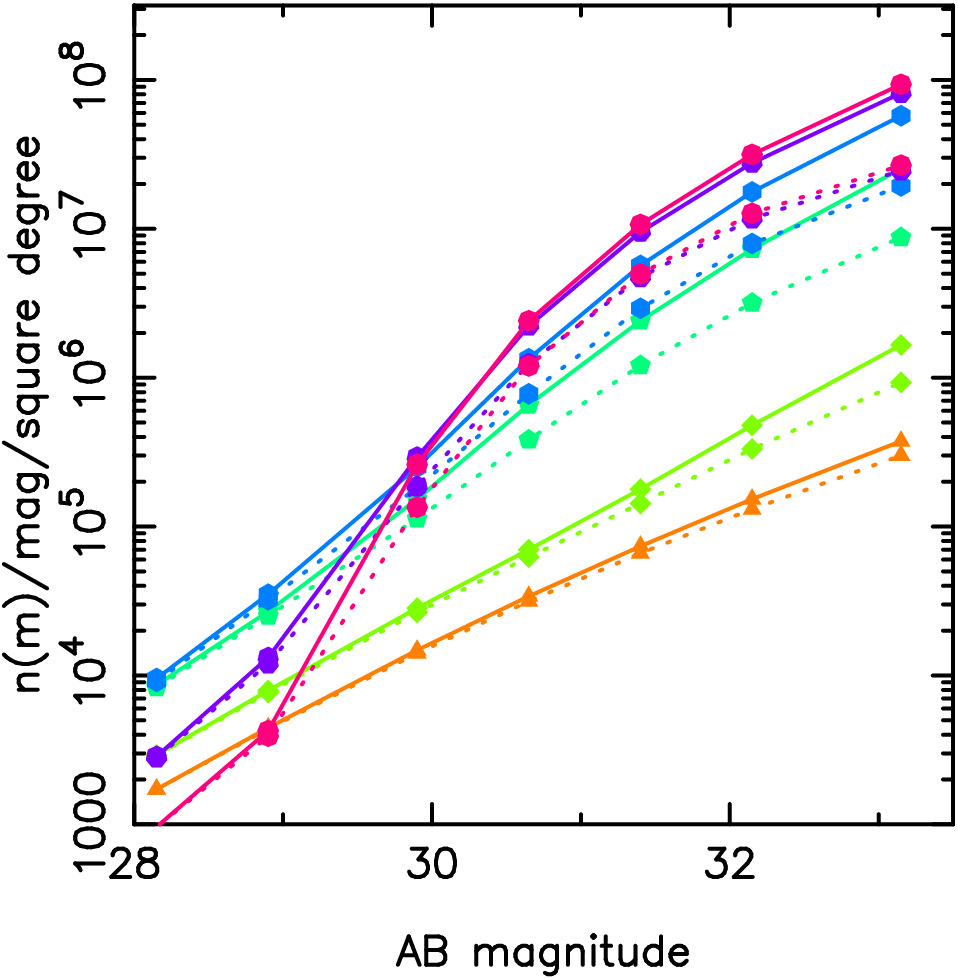]
{The number per magnitude per square degree as a function of limiting
AB magnitude for the V (triangles, orange), R (diamonds, light green),
J (pentagons, green), K (hexagons, blue), L (heptagons, purple) and M
(octagons, red) bands. The solid lines are evolving density models and
the dotted for fixed co-moving density models.
\label{fig:nm}}

\figcaption[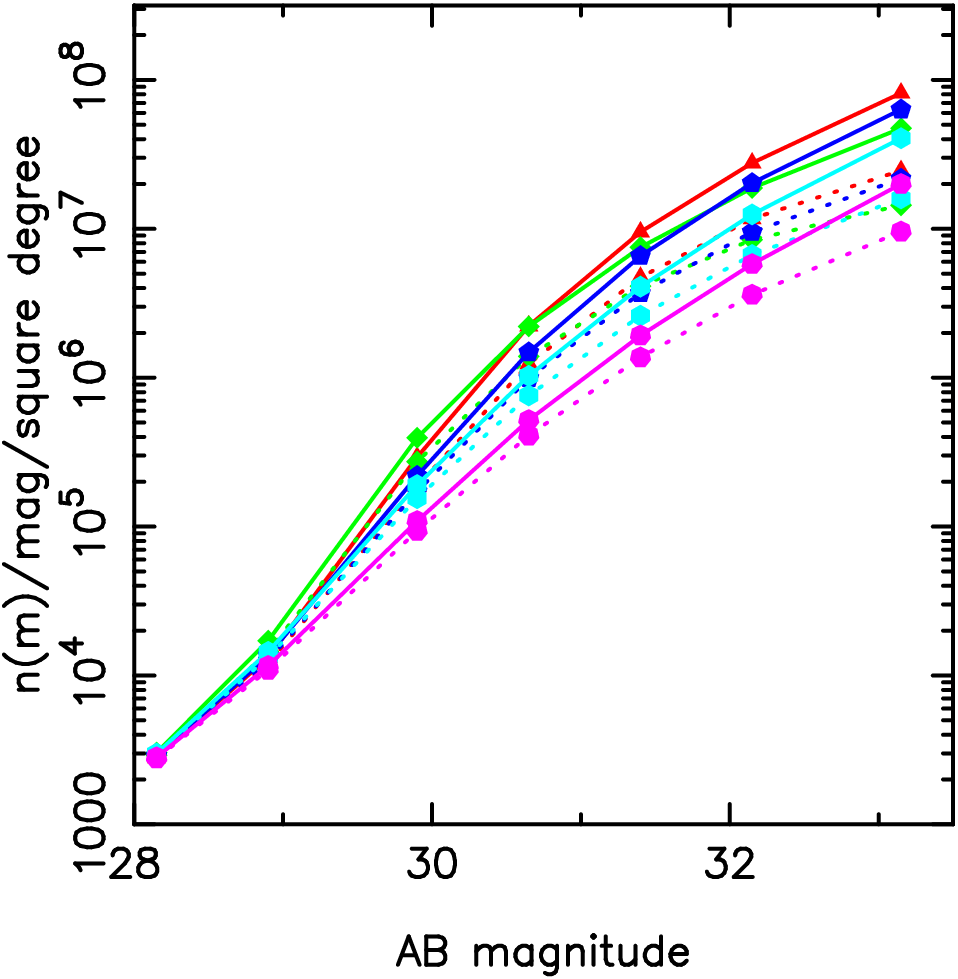]
{The L band counts for cluster formation in the 0-1 Gyr interval
(triangles, orange), the 1-2 Gyr interval (diamonds, light green), and
exponential models with $\tau=1$~Gyr (pentagons, green), 2~Gyr
(hexagons, blue) and 4~Gyr (heptagons, purple) for both evolving and
non-evolving mass functions.
\label{fig:nmE}}

\begin{figure} \figurenum{\nfig}\includegraphics{fig1.eps}\caption{}\end{figure}  
\begin{figure} \figurenum{\nfig}\includegraphics{fig2.eps}\caption{}\end{figure}  
\begin{figure} \figurenum{\nfig}\includegraphics{fig3.eps}\caption{}\end{figure}  
\begin{figure} \figurenum{\nfig}\includegraphics{fig4.eps}\caption{}\end{figure}  
\begin{figure} \figurenum{\nfig}\includegraphics{fig5.eps}\caption{}\end{figure}  

\end{document}